\documentclass[twocolumn,twoside]{article}
\usepackage{graphicx}
\usepackage{textcomp}
\usepackage{amsmath}

\title{Laser-controlled adaptive optic for beam quality enhancement in a multipass thin disk amplifier}

\author{Richard Lange\\
					Now: Physikalisch-Technische Bundesanstalt\\
					Bundesallee 100\\
					38116 Braunschweig\\
					\and
					Daniel Kolbe\\
					Deutsches Zentrum f\"ur Luft- und Raumfahrt e.V.\\ 
					Institut f\"ur Technische Physik, Pfaffenwaldring 38--40\\
					D-70569 Stuttgart}

\begin{document}

\date{1.6.2018}

\maketitle



\textbf{
We devise a laser-controlled adaptive optical element which operates intracavity under high intensity radiation. This element substitutes a conventional mechanically deformable mirror and is free of critical heat-sensitive components and electronics. The deformation mechanism is based on the projection of a CW control laser onto a specially designed mirror. Mounted to a water-cooled heat sink, the mirror can handle laser radiation beyond 3~MW/cm\textsuperscript{2}. The properties of the adaptive optical element including the maximum correctable wavefront pitch of 800~nm are discussed. The successful implementation in a multipass thin disk amplifier is presented. An improvement of the beam quality by a factor of three is achieved. We identify measures to enhance the performance of the adaptive optic towards efficient operation in a high-power laser system.
}

Adaptive optics (AO) have become an important tool for beam correction and shaping in laser systems, addressing a large variety of ongoing research topics. Considering the low-power regime, their application ranges from the medical sector for ophthalmoscopy \cite{Roorda.2002} to astrophysics for laser guide stars \cite{Liu.2018}. For high-power laser oscillators and amplifiers, thermally induced wavefront aberrations of the laser beam are a major challenge. Good beam quality can still be achieved through implementation of AO components in the setup withstanding high-intensity radiation. This enables applications of high-power laser systems which require excellent beam quality, such as laser effectors for space debris removal with an earth-based system \cite{Esmiller.2014}. A widely employed AO component is the deformable mirror (DM), commonly equipped with electronically addressed actuators on the backside to perform deformations mechanically. With the development of high-reflective coatings such mirrors have been successfully applied in high-power cw laser setups, allowing intensities up to 1.5~MW/cm\textsuperscript{2} \cite{Sinha.2002,Verpoort.2012}. 
The correction of wavefront aberrations and the enhancement of a laser beam's quality has been achieved with this type of mirror in serial \cite{Wittrock.2003} and parallel \cite{McNaught.2009} two-stage amplifiers. To correct for low-order aberrations, tip/tilt mirrors and lens systems may be used \cite{McNaught.2009,Yu.2017,Lai.2016}.
In an alternative implementation of the DM, specifically designed for low mechanical noise, resistors substitute the mechanical actuators to generate heat induced deformations \cite{Kasprzack.2013}.
Good results have also been demonstrated with a third, pneumatically adapted type of the deformable mirror. Inserted in the cavity of a thin disk laser this approach enables a near diffraction-limited beam quality of $M^2$~<~1.4 at pump powers of about 2~kW and output powers of up to 815~W \cite{Piehler.2012}. The drawback of the design is its limitation to the compensation of spherical wavefront distortions, thus higher order aberration terms can not be corrected for.  
\newline
In this paper we introduce a laser-controlled approach for the adaption of the DM, thermally addressing its surface by absorbing radiation of an intensity-modulated laser beam. To avoid confusion, we want to clearly distinguish between the control laser adapting the mirror and the primary laser being corrected by the AO. Instead of placing the actuators directly on the mirror, a generic spatial light modulator remotely modifies the AO surface via the control laser radiation. The removal of the actuators enables the installation of the mirror on a heat sink for operation of the AO at high primary laser power beyond 3~MW/cm\textsuperscript{2}. The AO is primarily designed for operation in a thin disk amplifier (TDA) \cite{Giesen.1994}, which consists of a thin laser active element placed on a heat sink. The TDA layout allows for efficient cooling of the active medium and high amplification power, but also suffers from heat induced deformations. The designs of our DM and the TDA are similar and the process causing wavefront deformations in the TDA is expected to be analog to the compensation mechanism of the DM, as both are heat induced processes. Thus our AO is expected to allow for an efficient correction of aberrations in a TDA. The AO is tested in a multipass thin disk amplifier (MTDA).

\begin{figure}[h]
\includegraphics[width = 0.45\textwidth]{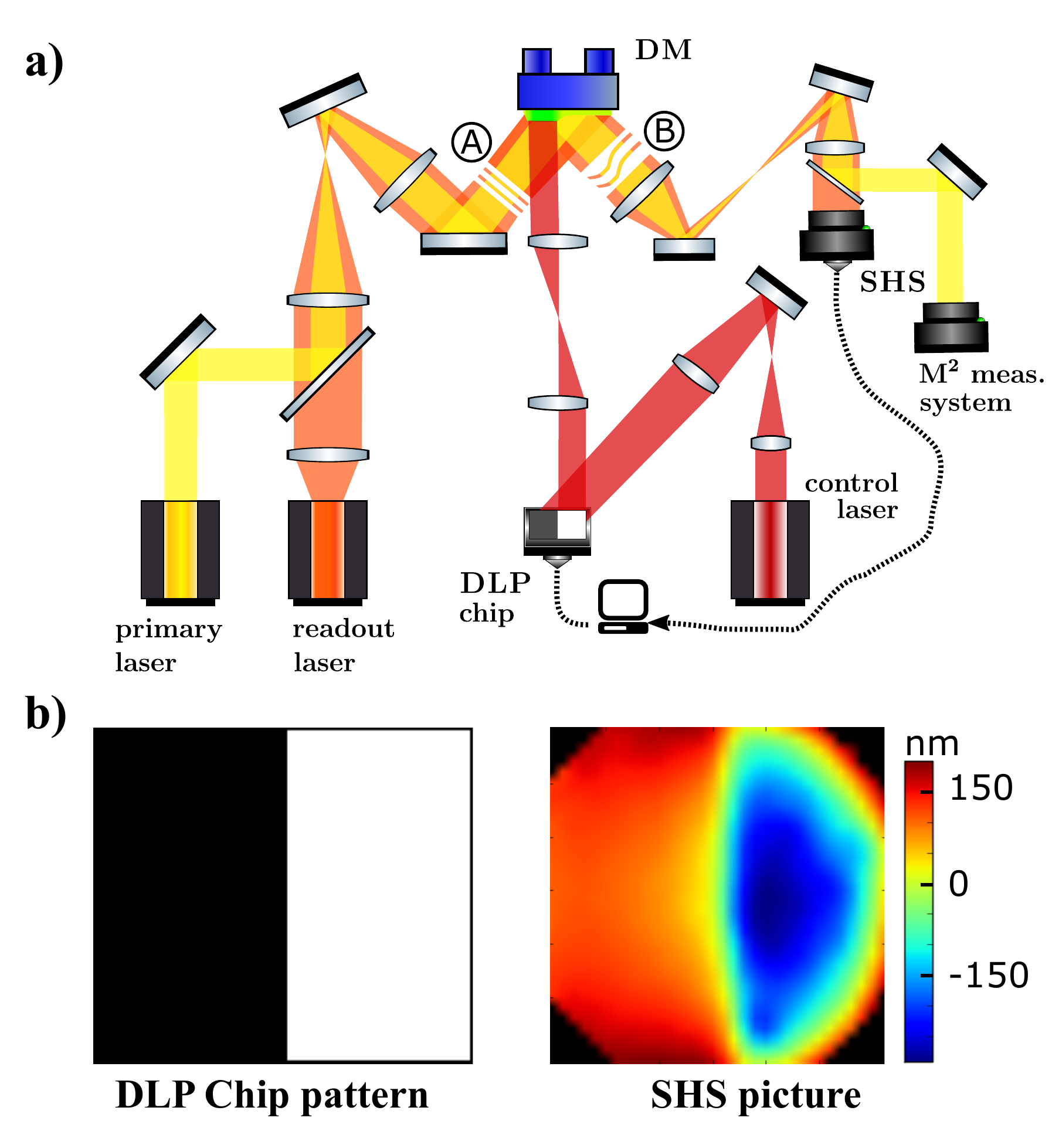}
\caption{a) Adaptive optic setup: The deformable mirror (DM), placed on a water-cooled heat sink, can be adressed with the control laser beam. The beam is intensity modulated by the DLP chip, exemplarily featuring the pattern shown in b). The readout laser beam is used to capture the mirror's deformation in its wavefront measured by the SHS. The aberrations are analyzed with a computer and results are fed back to the DLP chip, making the system closed loop controllable. Telescopes between the components ensure proper relay imaging. When implemented in a laser system, the primary laser whose wavefront needs to be corrected follows the readout laser path but is projected on a M\textsuperscript{2} measurement system for beam quality analysis. \newline b) Exemplary wavefront adaption: If a sharp edge is depicted on the DLP chip, a sharp cut-off in the control laser beam intensity leads to a defined plateau on the mirror surface. The wavefront of the readout laser is deformed as schematically illustrated with (A) and (B) in a) and recorded in the SHS picture.}
\label{graf: FIG1}
\end{figure}

\begin{figure}[h]
\centering
\includegraphics[width = 0.45\textwidth]{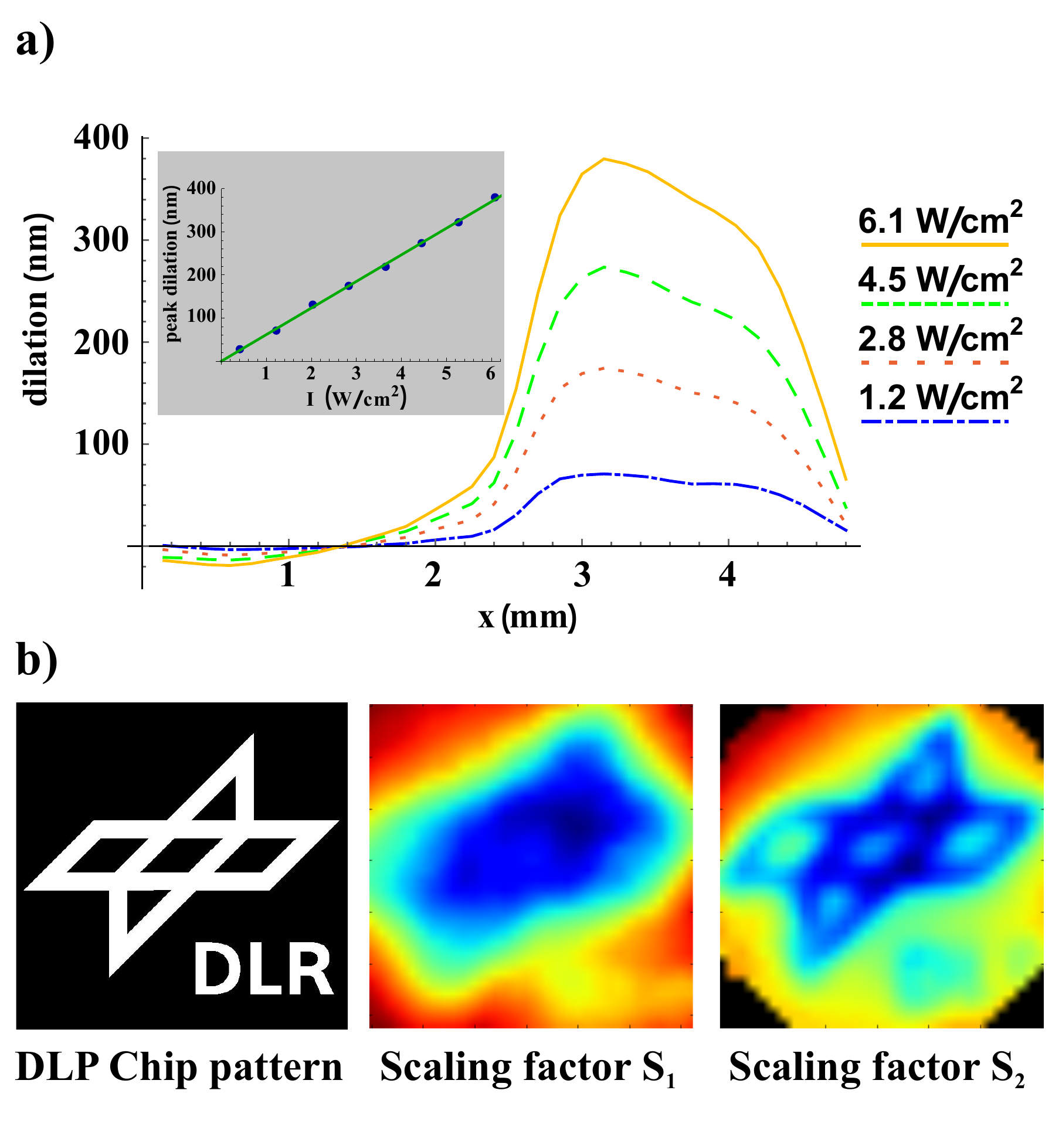}
\caption{a) Dilation of the mirror surface: Horizontal cut through the wavefront picture of the SHS with a sharp edge depicted on the DLP chip, as shown in Fig.~\ref{graf: FIG1}~b), at different intensities irradiated on the DM. The inlay illustrates the linear dependence between peak dilation and intensity on the AO. The decreasing dilation towards the edge results from non-uniform power distribution of the control laser beam and heat dissipation in the mirror material.
\newline
b) Resolution of the AO visualized with the DLR~Logo as the DLP chip pattern: The left picture shows the DLP chip pattern, which leads to different wavefront distortions in the other two pictures depending on the image scaling of the relay lens systems applied in our setup. In particular, switching from a 1:1~scaling between DM and SHS ($S_1$) to a 2:1~scaling ($S_2$) improves the resolution by a factor of two. As in the sharp edge pictures, a non-uniform dilation with respect to the original pattern is visible.}
\label{graf: FIG2}
\end{figure}

The adaptive optic shown in Fig.~\ref{graf: FIG1}~a) consists of three components: (1)~The deformable mirror is a 500~\textmu m thin, round Schott RG1000 filter with a diameter of 25~mm, absorbing below a wavelength of 1000~nm. It is provided with a highly reflective coating of 99.9~\% at 1030~nm wavelength (with 960~nm cut-on wavelength) and glued to a water-cooled copper heat sink. The mirror has been tested at radiation of 3~MW/cm\textsuperscript{2} without any damage. The coating is designed to withstand intensities up to 50~MW/cm\textsuperscript{2}.
(2)~The mirror is addressed using a diode laser (control laser) at 808~nm wavelength with an output power of 80~W whose beam is spatially intensity modulated with a Digital Light Processing (DLP) chip. The chip consists of a 1024x768 micromirror array that displays a grayscale picture, which is projected on the mirror's surface using a relay lens system. (3)~A second diode laser  (readout laser) at 980~nm wavelength is used to record wavefront deformations in the beam path (i.e. deformations of the mirror's surface and -- when implemented in the MTDA -- of the laser disk as well) that can be measured with a Shack-Hartmann Sensor (SHS). Results of the wavefront evaluation are analyzed with a computer and used to adjust the mirror via the DLP chip image, making the system closed loop controllable. Applying the AO system for the correction of wavefront deformations of a primary laser, the laser beam is coupled into the path of the readout laser beam. For the analysis of the beam quality, it is projected onto a M\textsuperscript{2}~measurement system after the AO.  
\newline
There are two critical parameters that determine the performance of an adaptive optic, its resolution and the maximally correctable wavefront pitch.
Fig.~\ref{graf: FIG2}~a) shows the mirror surface dilation for the sharp edge picture of Fig.~\ref{graf: FIG1}~b) at different intensities. A peak value of 400~nm enables corrections of distortions in the reflected wavefront up to 800~nm. The inlay shows the linear dependence on the irradiated intensity of approximately 62~nm/(W/cm\textsuperscript{2}). The maximal pitch is mainly limited by the intensity of the control laser at the DLP chip, which we restrict to 50~W/cm\textsuperscript{2} in order to prevent damage of the chip. Due to high losses of more than 75~\% at the DLP chip, caused for example by the coating, the fill factor and dead times in the control algorithm of the micromirror array, the intensity drops down to 6.1~W/cm\textsuperscript{2} at the DM. The dilation falls off to the edges due to a non-uniform power distribution in the control laser beam and due to heat dissipation to the unaddressed surface area of the DM. With a higher dynamic range of the irradiated intensity and a more sophisticated DLP chip pattern a smoother distribution of the dilation is expected, reproducing the flat plateau.
\newline
The second performance parameter, the resolution of the AO, can be determined from the measurement of Fig.~\ref{graf: FIG2}~a) as well. We define it as the width $d$ of the edge, from 10~\% to 90~\% of the maximal dilation. For the following discussion, we will distinguish more precisely between the resolution $d_{\text{m}}$ directly on the mirror and the resolution $d_{\text{wp}} = S \cdot d_{\text{m}}$ of the wavefront picture with an image scaling factor $S$. The resolution on the mirror $d_{\text{m}}$ is mainly limited by heat dissipation and mechanical stress between glass substrate and heat sink, it is independent of the irradiated intensity. For improvement, the manufacturing process of the DM needs to be revised. By increasing the image size between DLP chip and DM with a scaling factor $1/S ~ (S < 1)$, the adaptive area on the mirror is enlarged while $d_{\text{m}}$~remains unchanged: The relative resolution of the mirror is enhanced. The uncorrected wavefront is scaled with the same factor before the DM to match the enlarged adaptive area. Rescaling the corrected wavefront between DM and SHS with $S$ keeps the beam size constant within the setup but reduces $d_{\text{wp}}$ by $S$. An example is given in Fig.~\ref{graf: FIG2}~b) which shows imprints of the DLR~Logo onto the readout laser wavefront. We double the radius of the mirror's adaptive area, switch from a 1:1~image scaling between DM and SHS ($S_1 = 1$) to a 2:1~scaling ($S_2 = 1/2$) and enlarge the readout laser beam before the DM respectively. The resolution of the AO improves by a factor of two, with $d_{\text{wp}} = 1.7$~mm deduced from Fig.~\ref{graf: FIG2}~a) for the latter setup.
\newline

The setup of the multipass thin disk amplifier is based on the concept of the relay neutral gain module (NGM) as described in detail in \cite{Mende.2009,Mende.2012}. The primary laser at 1030~nm wavelength with a spotsize of 5.4~mm diameter and an input power of 1~W propagates through seven NGM, each consisting of an Yb:YAG laser disk (LD) for amplification and our DM for wavefront corrections in successive conjugated planes. Folded telescopes realized with plane and curved mirrors on a ringlike aluminum structure redirect the beam in all NGM on the same LD and DM. The respective relay lens systems generate a 1:1~image scaling between LD and DM, corresponding to a resolution as depicted in Fig.~\ref{graf: FIG2}~b), center. The LD is pumped at 940~nm wavelength with a pump spot diameter of 8~mm and a maximum power of 1750~W. After passing through the MTDA, the primary laser beam quality is measured with a camera on a translation stage according to ISO standard. The readout laser is coupled into the propagation path of the primary laser beam before the MTDA and directed on the SHS. The DM can be replaced by a plane mirror for removal of the AO from the setup.

In order to illustrate the successful application of the AO system in the MTDA, the quality of the amplified primary laser beam with and without the DM implemented in the setup has been measured and will be discussed in the following. The beam quality defined according to ISO~11146 has been chosen as the relevant factor since it comprises the key demands for a long-range high-power laser device, a high intensity at a large distance. The beam quality $M^2 = \pi/\lambda \cdot \omega_0 \cdot \theta_0$ for a given wavelength $\lambda$ restricts the smallest achievable beam waist $\omega_0$ in dependence of half the divergence angle $\theta_0$. A large $M^2$-value implies a large beam waist or divergence angle, thus at far distances the intensity, scaling inverse quadratically with the beam size, will strongly decrease.
\begin{figure}[h]
\centering
\includegraphics[width = 0.45\textwidth]{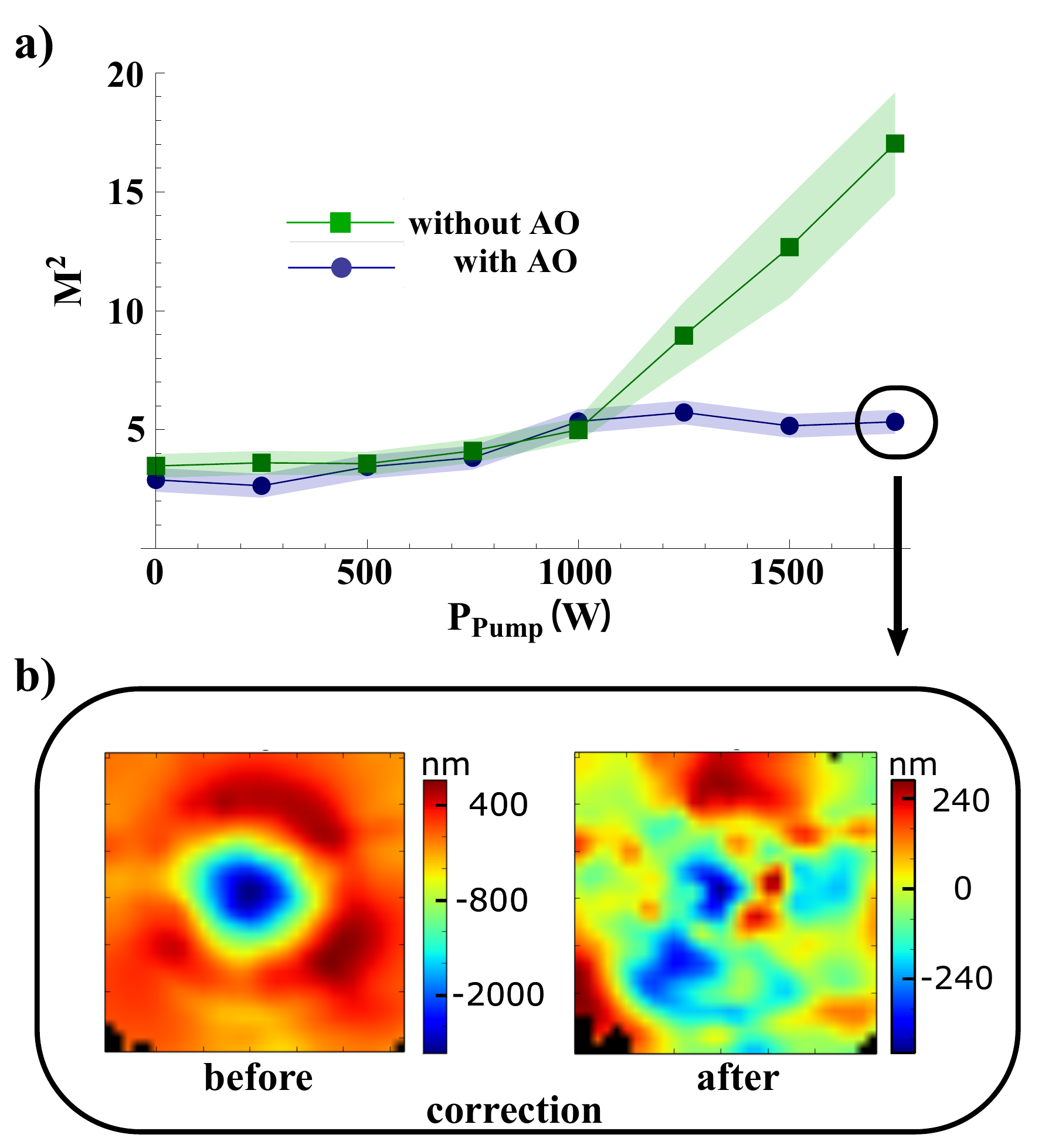}
\caption{a) Measurement of the beam quality $M^2$ as a function of the pump power P\textsubscript{Pump} with and without the AO implemented (i.e. the deformable or a plane mirror installed). At P\textsubscript{Pump}~=~1750~W, the beam quality can be enhanced by a factor of three. The shaded area depicts the uncertainty in the measurement. \newline b) For the measurement point at P\textsubscript{Pump}~=~1750~W with the AO installed, the wavefront pictures recorded with the SHS before and after the compensation process by the DM are shown. Tip-tilt and defocus terms in the aberration are removed as they do not affect the beam quality.}
\label{graf: FIG3}
\end{figure}
\newline
The beam quality has been measured on the two main axes of the beam for different pumping powers of the MTDA, with and without the AO implemented (i.e. with the deformable mirror or a plane mirror in the beam path installed). The results on both axes are similar, Fig.~\ref{graf: FIG3} a) shows their mean values. The shaded area depicts the uncertainty in the beam quality measurement, which is calculated from the fit of the $M^2$-formula on the beam caustic. At low pump powers, no change of the beam quality is measurable. In that power regime, the additional deformation due to the thermal load is negligible with respect to deformations of the laser disk itself. Increasing the pump power leads to a rapid deterioration of the beam quality if the AO system is not installed. However, with the DM implemented in the MTDA, the beam quality stays at a constant level of $M^2 = 5$ and is limited by the resolution of the deformable mirror. At the highest applied pump power of 1750~W an enhancement of the beam quality by a factor of three from $17 \pm 3$ to $5.4 \pm 0.5$ has been achieved. Regarding the performance of a long-range high-power laser system this corresponds to a threefold increase of the accessible propagation range at a predetermined intensity. For the respective measurement point, the wavefront deformations before and after compensation by the AO system are depicted in Fig.~\ref{graf: FIG3} b). Aberrations related to tip-tilt and defocus errors that do not affect the beam quality value are removed, they could be handled with lens and mirror systems. The wavefront is corrected up to the resolution limit of the deformable mirror. \newline
The heat induced deformations of the wavefront are expected to be stable over long periods of time as the laser is operating in a state of thermal equilibrium. The dynamics of the AO are therefore of secondary interest at the moment, but may be improved significantly in the future. The physical limit for the response time is set by the thermal expansion time constant $\tau$ of the DM with $\tau = (0.6 \pm 0.2)$~s. One iteration in the control loop takes about 8~s and about five iteration steps are needed for the loop to converge from the inactive AO to the optimal correction.  
\newline

The results regarding the improvement of an amplified laser beam's quality via compensation of wavefront aberrations with the AO system are very promising for the operation in a high-power MTDA. Necessary advancements include a reduction of the beam quality limit $M^2 = 5$ with the DM implemented. Presented experiments regarding the relay imaging between DLP-Chip and DM have shown that an adapted image scaling (i.e. switching from $S_1$ to $S_2$ scaling) will enhance the resolution of the AO system, thus allowing for wavefront corrections at a smaller scale, pushing the limit of the achievable beam quality to three or better. A higher dynamic range of the control laser intensity and further development of the DLP chip pattern for a smoother distribution of the mirror dilation will assist this process. Advances in the mirror geometry, e.g. a reduction of mechanical stress between glass substrate and heat sink, are expected to increase the resolution and thereby the beam quality even more. The response time of the AO is low, especially compared to an electronically adressed DM, but the dynamics do not pose an issue for our application. After implementing the discussed improvements, the system will be ready for the connection to a high-power primary laser beam to test the setup in a high-power MTDA.

A laser-controlled adaptive optic with the mirror deformed by absorption of an intensity-modulated laser beam has been introduced and its implementation in a multipass thin disk amplifier presented. Mounted to a water-cooled heat sink and free of heat-sensitive elements, the mirror withstands intensities exceeding 3~MW/cm\textsuperscript{2.} We have shown improvements in the quality of an amplified laser beam by a factor of three and identified measures to enhance the performance of our adaptive optic, paving the way for a successful application in high-power laser devices.
\newline

We thank Elke Schmid for preliminary work, and Daniel Sauder and Thomas Dekorsy for valuable discussions and critical reading of the manuscript.

\bibliographystyle{unsrt}
\bibliography{bibliography}

\end{document}